\documentclass[oldversion]{aa}
 \usepackage{txfonts} \usepackage{epsfig} \usepackage{graphicx}
 \usepackage{natbib} \bibpunct{(}{)}{;}{a}{}{,}

 \newcommand{\kms}{{\km\second^{-1}}}
 \newcommand{\solarmass}{{\,{\textnormal{M}}_\odot}}
 \newcommand{\AU}{{\,\textnormal{AU}}}
 \newcommand{\km}{\,{\textnormal{km}}}
 \newcommand{\second}{\,{\textnormal{s}}}
 \newcommand{\stellarmass}{{{\textnormal{M}}_*}}

\begin{document}

 \titlerunning{A rotating molecular jet in Orion}

 \authorrunning{Zapata et al.}

 \title{A rotating molecular jet in Orion}

 \author{Luis A. Zapata, Johannes Schmid-Burgk, Dirk Muders, Peter
 Schilke, Karl Menten, and Rolf Guesten}

 \institute{Max-Planck-Institut f\"{u}r Radioastronomie, Auf dem
 H\"ugel 69, 53121, Bonn, Germany}

 \date{Received -- / Accepted --}

 \offprints{Luis Zapata, \email{lzapata@mpifr-bonn.mpg.de}}

 \abstract{We present  CO(2-1), $^{13}$CO(2-1),
CO(6-5), CO(7-6), and
 SO(6$_5$-5$_4$) line observations made with the {\it IRAM 30 m} and Atacama 
Pathfinder Experiment ({\it APEX}) radiotelescopes
 and the {\it Submillimeter Array (SMA)} toward the highly collimated
 (11$^{\circ}$) and extended ($\sim$ 2$'$) southwest lobe of the
 bipolar outflow {\it Ori-S6} located in the Orion South region.  
 We report, for all these lines, the detection of 
 velocity asymmetries about the flow axis, with velocity differences 
roughly on the order  of
1 km s$^{-1}$
 over distances of about 5000 AU, 4 km s$^{-1}$
 over distances of about 2000 AU, and close to the source of between 
 7 and 11 km s$^{-1}$ over smaller scales of about 1000 AU.
 The redshifted gas velocities are located to the southeast of the
 outflow's axis, the blueshifted ones to the northwest.  We interpret
 these velocity differences as a signature of rotation but also
 discuss some alternatives
 which we recognize as unlikely in view of the asymmetries' large downstream
 continuation.  In particular, any straightforward interpretation
by an ambient velocity gradient does not seem viable.
This rotation across the {\it Ori-S6} outflow is
 observed out to (projected) distances beyond 2.5 $\times$ 10$^4$ AU
 from the flow's presumed origin.  Comparison of our large-scale (single dish) and
 small-scale (SMA) observations suggests the rotational velocity to
 decline not faster than 1/R with distance R from the axis; in the innermost
few arcsecs an increase of rotational velocity with R
is even indicated. The magnetic field
lines threading the inner rotating CO shell may well be anchored in a disk
of radius $\sim$ 50 AU; the field lines further out need a more extended
rotating base.
 Our high angular resolution SMA observations also suggest
 this outflow to be energized by the compact millimeter radio source
 139-409, a circumbinary flattened ring that is located in a small
 cluster of very young stars associated with the extended and bright
 source FIR4.}

 \maketitle

 \section{Introduction}

Protostellar jets have the essential task of removing angular momentum
from the cores of pre-/protostellar clouds, in order for these to
contract into new stellar objects.  It is believed that T Tauri winds
and protostellar jets are driven magnetocentrifugally from keplerian
accretion disks close to the central stars (for a review see
\citet{Konigletal2000,Shuetal2000,Pudritzetal2007,Shangetal2007}).  In
these models the magnetic fields that are anchored in the accretion
disk-star system are responsible for accelerating the jets' material
from the accretion disk.  The material ejected from the disk therefore
possesses angular momentum and, if not completely free to move away
from the jet immediately, will thus show a toroidal velocity
component, {\it i.e.} rotation about the jet axis.

 In recent years a number of observations at optical wavelengths,
 using high spectral and angular resolution toward the launching zones
 of young jets from T Tauri stars, have attempted to identify velocity
 asymmetries that might be interpreted as signatures of jet rotation.
 These observations include \citet{Bacciottietal2002},
 \citet{Coffeyetal2004}, \citet{Woitasetal2005}, and
 \citet{Coffeyetal2007} who detected systematic velocity shifts (of
 order 5 to 25 km s$^{-1}$) across the jet's axis within the first
 100-200 AU from the star, toward six T Tauri stars, {\it e. g.} DG
 tau, RW Aur, CW Tau, and Th 28. Furthermore, observations at infrared
 wavelengths have also revealed such velocity jumps in the
 launching zones of the Herbig-Haro objects HH 26 and HH 72
 \citep{Chrysostomouetal2008}.  All these authors have interpreted the
 velocity shifts as a signature of jet rotation produced by a
 magneto centrifugal wind.

 The first tentative evidence of such an outflow rotation was
 presented by \citet{Davisetal2000} at infrared wavelengths. They
 reported velocity shifts of a few km s$^{-1}$ across the HH212 jet at
 distances of about 10$^4$ AU from the ejecting object using spectral
 line observations of the molecule H$_2$.

 At millimeter wavelengths there have likewise been attempts to find
 such signatures near the base of strong molecular outflows (HH 30:
 \citet{Petyetal2006}; HH212: \citet{Codellaetal2007}, \citet{Leeetal2008}; 
 HH 211: Lee et al. (2007); CB 26: \citet{lau2009}).  
 However, only in the CB 26 and the HH212 outflows there seems to be evidence of rotation.  

\begin{figure*}[ht]
\begin{center}
\includegraphics[scale=0.55]{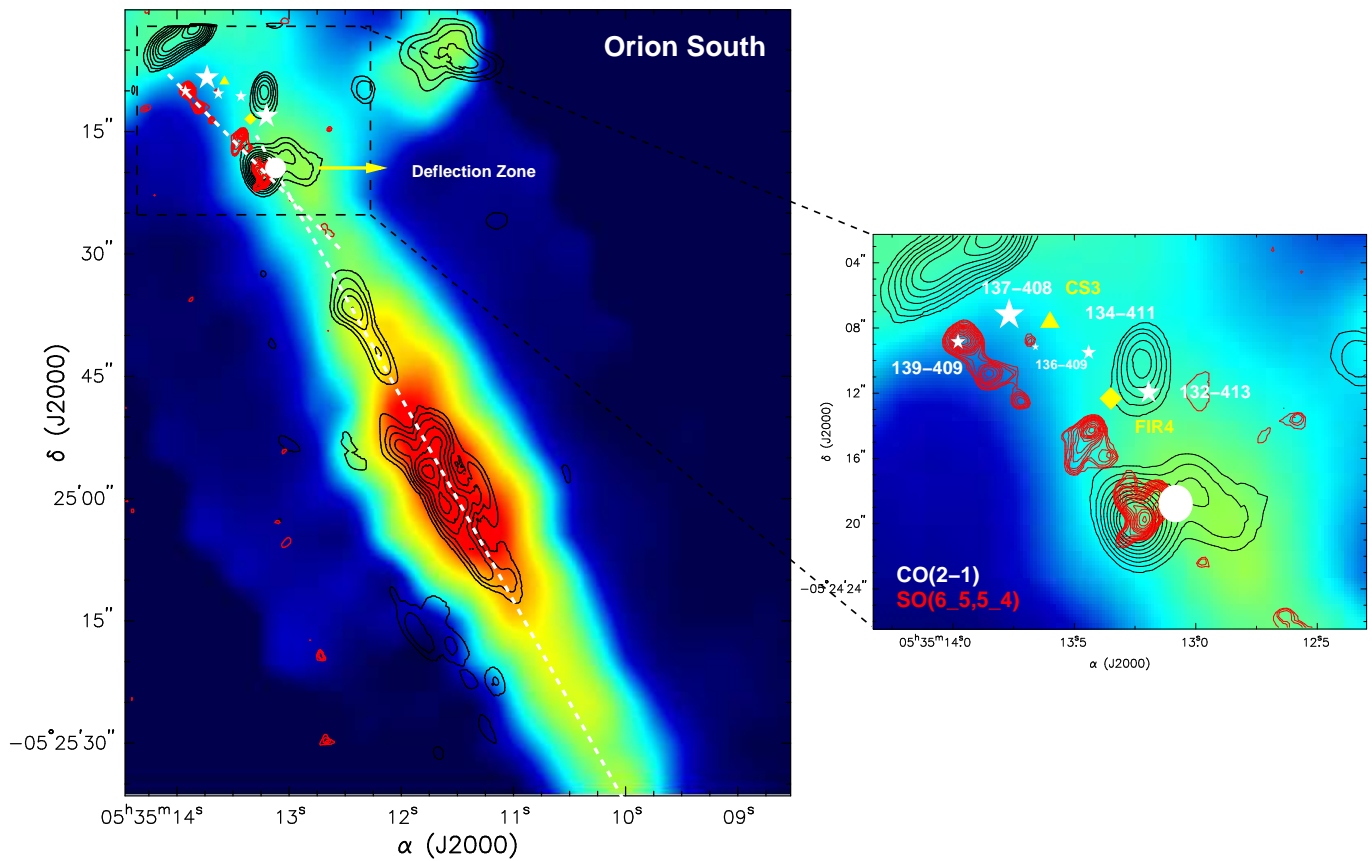}
\caption{\scriptsize  IRAM CO(2-1) moment zero map of the {\it Ori-S6} 
 collimated molecular outflow ({\it color image}) overlayed
 with the SMA moment zero of the CO(2-1) ({\it black contours}) and SO(6$_5$-5$_4$) 
 ({\it red contours}) images.
 For CO the contours are 4, 6, 8, 10, 12, 14, 16, 18, 
 and 20 times 1.1 Jy beam$^{-1}$ km s$^{-1}$, the rms-noise of the image, while for 
 SO the contours are 4, 5, 6, 7, 8, 9, 10, 11, 12, 13, 14, 15, 16, 17, 18, 19, and 20
 times 0.7 Jy beam$^{-1}$ km s$^{-1}$, the rms-noise of the image.
 The molecular emission in the SO map is integrated over velocities between $+$ 8
 and $+$ 25 km s$^{-1}$, for the SMA CO(2-1) map from $+$ 11 to $+$ 25 km s$^{-1}$ and for the
 IRAM 30m CO(2-1) map from $+$ 12 to $+$ 23 km s$^{-1}$.
 The yellow rhombus and triangle denote the position of the source FIR 4 (Mezger et al. 1990) 
 and the millimeter source CS 3 (Mundy et al. 1986), respectively. The
 white stars mark the position of the compact millimeter sources reported by
 \citet{Zapataetal2005}. Note the SO(6$_5$-5$_4$) maps did not cover the same large areas
 as the CO(2-1) maps from IRAM 30m and SMA. The dashed white lines mark approximately  
 the axis of the outflow before and after deflection. The white dot marks the point O.
 }
\label{fig1}
\end{center}
\end{figure*}

 Recent studies, however, have proposed alternative explanations for
 observed velocity asymmetries in the jets. \citet{Sokeretal2005}
 remarked that the interaction of the jet with a twisted-tilted
 (warped) accretion disk can lead to the observed asymmetry in the
 jet's line-of-sight velocity profile and thus the magneto centrifugal
 wind acceleration model is not required to explain such velocity
 jumps at the base of the optical and infrared
 jets. \citet{Cerqueiraetal2006} proposed that a precessing jet whose
 ejection velocity changes periodically with a period equal to the
 precession period could also reproduce the line profiles of the jets.

In the following we address the issue of outflow rotation by combining
new interferometer and single-dish observations of the {\it Ori-S6}
molecular outflow \citep{Schmid-Burgketal1990} which is located in
the southernmost part of the very active high- and intermediate-mass
star forming region Orion South. 
Single-dish observations suggested this outflow to originate near the
millimeter sources CS3/FIR4 some 100$''$ to the South of the BN/KL
object. Unfortunately the immediate vicinity of CS3/FIR4 contains at least 
two more molecular outflows \citep{Zapataetal2005,Zapataetal2007}, both almost perpendicular to
Ori-S6, so that unique differentiation between the several emitting
structures becomes possible for single telescopes only beyond some 20$''$
from the presumed region of origin. Interferometry however has permitted
\citet{Zapataetal2004a,Zapataetal2004b} to propose the source
of this outflow to be the elongated radio object 134-411 (possibly a
compact thermal jet) that is located close to the position of
CS3/FIR4, the position angle of its major axis being consistent with
that of the outflow. 

The redshifted lobe of Ori-S6 is
highly collimated (11$^{\circ}$), of relatively low radial velocities (up to
15 to 20 km s$^{-1}$ relative to ambient) and extended over $\sim$
2$'$ in a direction of P.A. around 210$^{\circ}$.  That lobe is much
better defined and stronger than its blueshifted counterpart because
it is interacting directly with the molecular cloud, while the latter
points in direction of the HII region and is likely destroyed by it.
In the following we will deal with the redshifted lobe only.

\citet{Zapataetal2006} suggested that this outflow
could be associated with a collimated SiO jet that they had detected
in their SMA observations and that is emanating from the CS3/FIR4
region with almost the same orientation. On larger scales, \citet{Henneyetal2007}
discussed the possibility that the redshifted lobe of the {\it Ori-S6}
outflow is associated with a strong shock ("Southwest Shock"), some
400$''$ downstream, that is observed at optical wavelengths and imaged
in their Fig. 6.  

CO position-velocity diagrams along the red lobe reveal a two-step structure
over a distance of about 120$''$ from the source region. Over the inner 60$''$
there is a linear buildup of outflow velocities, from ambient (around 7 to 8 km s$^{-1}$) 
to maximum values near 25 km s$^{-1}$,
which then terminates abruptly. At the same
point the ambient velocities as seen in CO, which closer to CS3/FIR4 cover a
rather wide range of between 7 and 9.5 km s$^{-1}$, acquire an additional component,
of around 10 km s$^{-1}$, which further downstream becomes the dominant velocity.
At this point a second, even faster linear buildup of radial velocity, again
starting from ambient values, sets in which follows exactly the same direction as
the first one. Quite likely the break in outflow velocity around the 60$''$
point is related to the change in ambient conditions.

\begin{figure*}[ht]
\begin{center}
\includegraphics[scale=0.35]{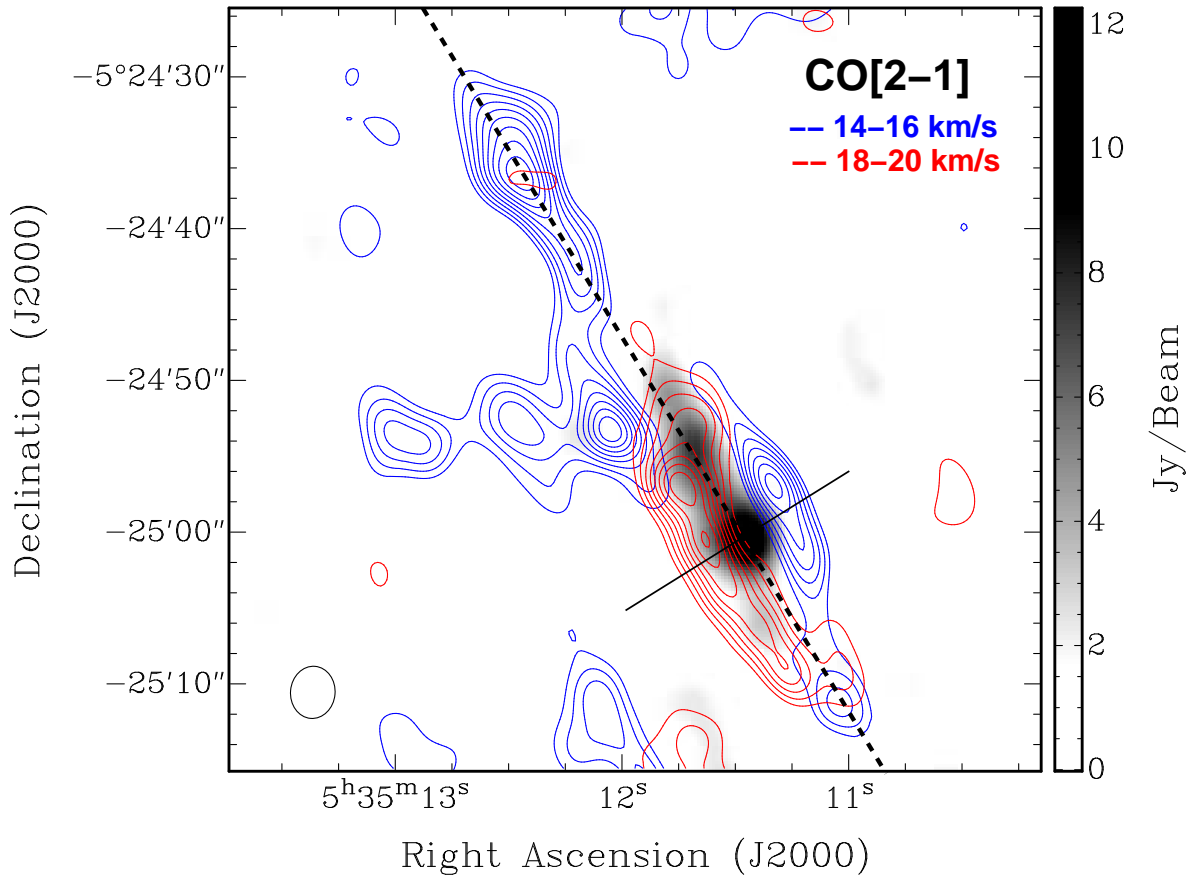}
\caption{\scriptsize Moment zero CO(2-1) contour map of the S6-Ori outflow showing the
rotating shell (blue and red contours) and the inner high velocity jet (grey scale).  
 The contours are from 15\% to 90\% with steps of 10\% of the peak of the line emission, 
the peak being 11 Jy beam$^{-1}$. The redshifted emission 
({\it red contours)} is integrated from 14 to 16 km s$^{-1}$, the blueshifted one ({\it blue contours)} 
from 18 to 20 km s$^{-1}$, and the grey one from 20 to 30 km s$^{-1}$. Note how the redshifted component
also traces low velocity gas from the jet toward the west. The dashed line marks approximately the 
outflow's symmetry axis and the continuous black line traces the position where the position-velocity diagram
in Figure 3 was made. The blueshifted emission to the east of the outflow located along about 
-24$'$ 50$''$ is an artifact from the interferometric image.
                     The synthesized beam is
                     shown in the bottom left corner of the image and
                     has a size of $3\rlap.{''}44$ $\times$ $2\rlap.{''}93$ with a P.A. =-4.9$^\circ$. 
                     Note that the outflow accelerates from top to bottom. 
                     }
\label{fig2}
\end{center}
\end{figure*}

In the present study we will concentrate on the inner 60$''$ section. Already early on
\citet{Murdersetal1992} had noted this section, as seen in C$^{18}$O(2-1), 
to be enveloped by a somewhat symmetrical cylindrical
structure that seemed to undergo rotation about the outflow axis.

 In this paper we present  CO(2-1), $^{13}$CO(2-1), CO(6-5), CO(7-6),
and SO(6$_5$-5$_4$) line observations made with both the IRAM 30 m
and APEX
 telescopes as well as the Submillimeter Array toward the redshifted lobe of {\it
 Ori-S6}.  We report the detection, at all three observatories,
of velocity jumps across the
 flow axis and interpret these as a signature of rotation. We discuss
 some alternative suggestions to explain these velocity asymmetries.
 Finally, the SMA observations suggest the source of the
 extended collimated outflow to be the millimeter continuum source
 139-409.

 \section{Observations}

 \subsection{Submillimeter Array}

 \subsubsection{SO(6$_5$-5$_4$)}

 Observations were made with the Submillimeter Array
 (SMA)\footnote{The Submillimeter Array is a joint project between the
 Smithsonian Astrophysical Observatory and the Academia Sinica
 Institute of Astronomy and Astrophysics, and is funded by the
 Smithsonian Institution and the Academia Sinica.}  during 2004
 September 3. The SMA was in its extended configuration, which
 includes 21 independent baselines ranging in projected length from 16
 to 180 m. The phase reference center of the observations was R.A. =
 05$^h$35$^m$14$^s$, decl.= -05$^\circ$24$'$00$''$ (J2000.0).  The
 size of the primary beam response at this frequency is 50$''$.  The
 receivers were tuned to a frequency of 230.534 GHz in the upper
 sideband (USB), while the lower sideband (LSB) was centered on
 220.534 GHz.

 The SO(6$_5$-5$_4$) transition was detected in the LSB at a frequency
 of 219.949 GHz. The full bandwidth of the SMA correlator is 4 GHz (2
 GHz in each band).  The SMA digital correlator was configured in 24
 spectral windows (``chunk'') of 104 MHz each, with 32 channels
 distributed over each spectral window, providing a resolution of 3.25
 MHz (4.29 km s$^{-1}$) per channel.

 The zenith opacity ($\tau_{230 GHz}$), measured with the NRAO tipping
 radiometer located at the Caltech Submillimeter Observatory, was
 $\sim$ 0.09, indicating very good weather conditions during the
 experiment.  Observations of Callisto provided the absolute scale for
 the flux density calibration.  Phase and amplitude calibrators were
 the quasars 0423-013 and 3C 120, with measured flux densities of 2.30
 $\pm$ 0.06 and 0.50 $\pm$ 0.03 Jy, respectively.  The absolute flux
 density calibration uncertainty is estimated to be 20\%, based on SMA
 monitoring of quasars.  Further technical descriptions of the SMA and
 its calibration schemes can be found in \citet{Hoetal2004}.

 The data were calibrated using the IDL superset MIR, originally
 developed for the Owens Valley Radio Observatory
 \citep{Scovilleetal1993} and adapted for the SMA.\footnote{The MIR
 cookbook by C.  Qi can be found at
 http://cfa-www.harvard.edu/$\sim$cqi/mircook.html} The calibrated
 data were imaged and analyzed in standard manner using the MIRIAD and
 AIPS packages.  We used the ROBUST parameter set to 0 for an optimal
 compromise between sensitivity and angular resolution.  The line
 image rms-noise was 20 mJy beam$^{-1}$ for each channel at an angular
 resolution of $1\rlap.{''}13$ $\times$ $0\rlap.{''}93$ with a P.A. =
 -73$^\circ$.

\begin{figure}
\begin{center}
\includegraphics[scale=0.35, angle=0]{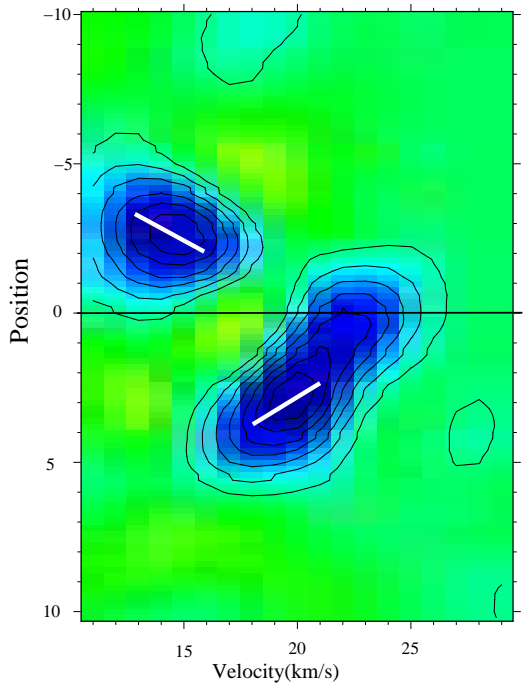}
\caption{\scriptsize Position-velocity diagram of the transversal cut 
 (along P.A. 125$^\circ$), marked in Fig. 2, across the outflow's redshifted bowshock. 
 The contours are 30\%, 40\%, 50\%, 60\%, 70\%, 80\%, and 90\% of the peak 
 of the line emission, the peak being 3.50 Jy beam$^{-1}$. The systemic LSR velocity
 of the ambient molecular cloud here is about 7 to 8 km s$^{-1}$. The black
 line marks the position of the symmetry axis of the
 molecular bowshock. The velocity and spatial resolutions are 1 km s$^{-1}$ and $\sim$ 3$''$, respectively. 
 The spatial scale is in arcsec. 
 For the white lines see text. }
\label{fig3}
\end{center}
\end{figure}

 Observations were made during December 2007. 
 The SMA was in its compact configuration, which
 includes 28 independent baselines ranging in projected length from 8
 to 57 m.  The receivers were tuned to a frequency of 230.534 GHz 
 in the upper sideband (USB), while the lower sideband (LSB) was 
 centered at 220.534 GHz. The observations were made in mosacing mode 
 using the half-power point spacing between field centers and thus covering 
 a total area of about 2$'$ $\times$ 2$'$ in Orion South that contains the
 outflow reported by \citet{Schmid-Burgketal1990}.

 The full bandwidth of the SMA correlator is 4 GHz.  
 The SMA digital correlator was configured in 24
 spectral windows of 104 MHz each, with 128 channels
 distributed over each spectral window, providing a resolution of 0.812
 MHz (1.05 km s$^{-1}$) per channel.

 The zenith opacity ($\tau_{230 GHz}$), was
 $\sim$ 0.1 -- 0.3, indicating reasonable weather conditions.  
 Observations of Uranus provided the absolute scale for
 the flux density calibration.  Phase and amplitude calibrators were
 the quasars 0530+135 and 0541-056.  

The line image rms-noise was 150 mJy beam$^{-1}$ for each channel at an angular
resolution of $3\rlap.{''}44$ $\times$ $2\rlap.{''}93$ with a P.A. =
-4.9$^\circ$.

\begin{figure}
\begin{center}
\includegraphics[scale=.75, angle=-90]{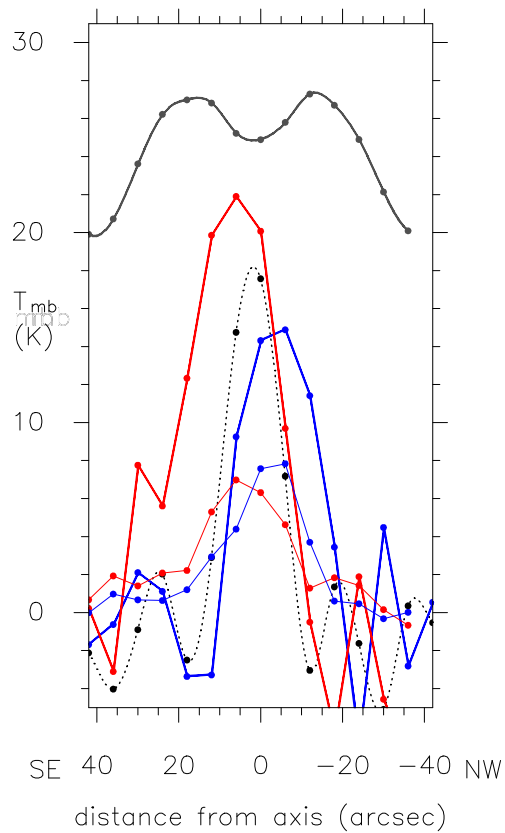}\vspace{2cm}
\caption{\scriptsize  Variation of intensities (IRAM 30m) 
across the outflow at downstream distance 45$''$ from point O:
Top curve is $^{13}$CO(2-1) at 7.7 km/s (i.e. ambient), shifted upwards by 8 K;
dotted line the outflow at high velocities, i.e. around 15 km/s. Strong (CO(2-1))
and weak ($^{13}$CO(2-1) $\times$ 7) color lines 
demonstrate the spatial separation between redshifted (11.6 km/s) and blueshifted
(4.6 km/s) components.  }
\label{fig4}
\end{center}
\end{figure}
\subsubsection{$^{12}$CO(2-1)}

\begin{figure*}[!ht]
\begin{center}
\includegraphics[scale=0.9, angle=-90]{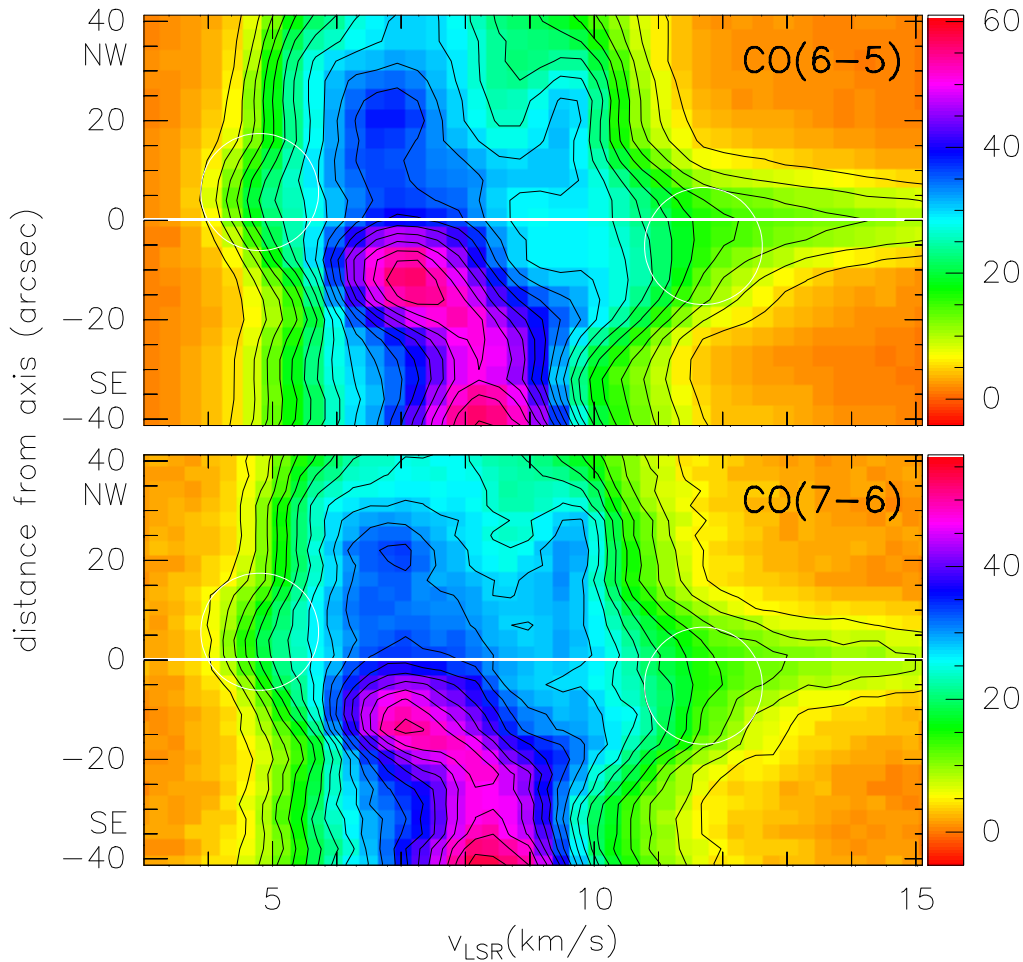}
\caption{\scriptsize Position-velocity diagrams (CO(6-5) and CO(7-6)) of the perpendicular crosscut through the
outflow at downstream distance 37$''$ from point O. The outflow
axis is marked by the horizontal white line, the regions discussed in the text for velocity
asymmetry is indicated by the circles. Colour scale is $T_A^*$ in K. }
\label{fig5}
\end{center}
\end{figure*}

 \subsection{Single-dish Observations}

 \subsubsection{ \it IRAM 30m}

Initial CO and $^{13}$CO(2-1) observations were performed in February 1991 at the
IRAM 30m telescope, covering a 70$''$ $\times$ 80$''$ area centered 10$''$
north of the presumed source of the outflow. The beam size at 220 GHz was
13$''$, the main beam efficiency 0.46. We used position switching with
the reference fixed at 30$'$ west of the grid center and chose a
velocity resolution of 0.14 km s$^{-1}$.  Observed points were 6$''$
apart, and the observation time per point was 80 s.

These measurements were extended and complemented by  $^{13}$CO(2-1)
OTF observations on February 2 and 4, 2008, with a
total observing time per night of 45 min. Now the center of the
100$''$ $\times$ 72$''$ grid was placed 36$''$ downstream from the
presumed source and the grid rotated to align with the outflow.  We co-added the
OTF dumps every 6$''$ along scans 6$''$ apart. Velocity resolution was
0.05 km s$^{-1}$. Both frequency switching (throw 15 MHz) and position switching
were used.

{ \subsubsection{ \it APEX}

Some preliminary CO(6-5) and (7-6) measurements were performed in October 2007
with the 12m Atacama Pathfinder Experiment (APEX) telescope located on Chajnantor in Chile. 
The same lines were
simultaneously reobserved in depth on September 21 and 22, 2008 with the 2 $\times$ 7-element
CHAMP$^+$ array, covering a 120$''$ $\times$ 120$''$ field around the origin of Ori-S6. Zenith opacity 
at 5100 m altitude was $\sim$ 0.35 at 690 GHz and 0.42 at 806 GHz.
At 690 GHz the APEX beam is 9$''$, at 806 GHz 7.5$''$. Spectral resolution was
chosen to be 0.73 MHz, corresponding to 0.32 resp. 0.27 km s$^{-1}$ per channel, such that
the 2048 channels per band resulted in a total velocity coverage of 650 resp. 550 km s$^{-1}$
per band, sufficient to detect the highest-velocity bullets possibly present in this 
very active star-forming region Orion South.

The single-dish measurements were reduced with the standard CLASS software of
the Gildas package\footnote{http://www.iram.fr/IRAMFR/GILDAS}, with only
first-order baselines subtracted from the data.

\begin{figure*}[!ht]
\begin{center}
\includegraphics[scale=1, angle=-90]{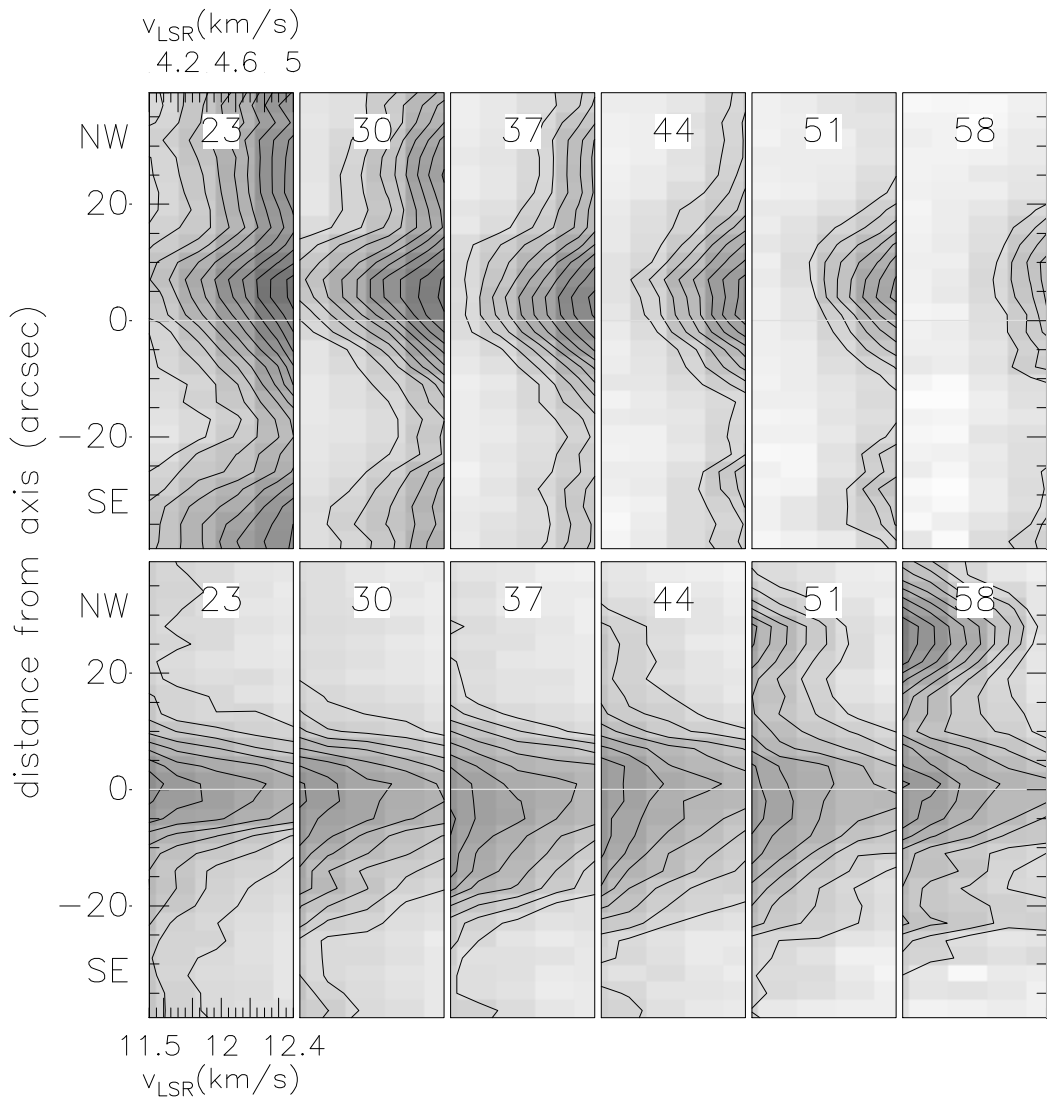}
\caption{\scriptsize Sections of the position-velocity diagrams (CO(7-6)) of perpendicular crosscuts through the
outflow at different distances (given by the numbers on the panels, arcsec) from point O. The outflow
axis is marked by the horizontal grey line.
Top: Blueshifted velocities between 4 and 5 km s$^{-1}$, bottom: Redshifted ones from
11.5 to 12.5 km s$^{-1}$.
Note that the structures near the outflow axis persist for at least six beams. }
\label{fig6}
\end{center}
\end{figure*}

 \section{Results}

 Fig. \ref{fig1} shows a map of the high velocity CO(2-1)
 emission from the Ori-S6 redshifted lobe, {\it i. e.} intensity
 integrated from $+$ 12 to $+$ 23 km s$^{-1}$, as obtained at the IRAM
 30 m telescope (Schmid-Burgk et al. 1990), overlaid with both our SMA
 total moment-zero SO(6$_5$-5$_4$) and CO(2-1) maps.  
 In this image one can see the SO emission to trace the
 inner, very collimated jet ejected along a P.A. of 45$^\circ$ from
 the source 139-409 that is located in a small cluster of young stars
 associated with the extended and bright source FIR4
 \citep{Zapataetal2006, Zapataetal2007}.  The CO emission on the other
 hand delineates the more extended parts of the outflow, with its
 major axis bending, within a small distance from the source, towards
 a P.A. of some 30$^\circ$. The small difference between both position
 angles can be explained by the outflow undergoing a smooth
 deflection, possibly due to the high density ($\sim$ 10$^6$ cm$^3$)
 molecular cloud located behind the Orion Nebula as had already been
 suggested for other outflows populating this region
 \citep{Zapataetal2006, Henneyetal2007}.  A process for deflection of 
 an outflow has been proposed and modeled by Cant\'o \& Raga (1996).
 But there is also the possibility that we are seeing two 
 outflows with different orientations, one traced by SO(6$_5$-5$_4$) on 
 small scales, and the other one by 
CO(2-1) on large scales. However, we do not see
 any CO evidence of an outflow with P.A.= 45$^\circ$ extending from near the source 
 FIR4. Moreover, coincidentally both components show the same velocity range of between 
 $+$8 and $+$25 km s$^{-1}$, indicating them to be part of the same outflow. 
 Furthermore, the SO clump or bullet named D in Fig. 1 is 
 well coincident with a bright clump of the CO outflow located in the 
 deflection zone, thus linking directly the
two components. Finally, Zapata et al. (2009, in prep.) found the continuation of 
the SO emission at the same position of the CO(2-1) shell, confirming thus 
that both molecular outflows are part of the same one. 
    
 Fig. \ref{fig2} presents an overlay of two CO(2-1) emission intervals of 
the south-western lobe of the Ori-S6 outflow made with the SMA, one integrated
over the velocities from 14 to 16 km s$^{-1}$ (blue), the other from 18 to 20
km s$^{-1}$ (red). This shows clearly 
a velocity ``jump'' across the outflow, with the redshifted gas velocities located toward 
the South-east, the blueshifted ones toward
the North-west. With a distance of 415 pc to the Orion Nebula \citep{Mentenetal2007}, 
the separation between these two components       
is about 2 $\times$ 10$^3$ AU.

The position-velocity diagram of Fig. 3, taken along the black line across the
outflow in Fig. 2, displays the decrease of radial velocity in the rotating shell
with distance from the flow axis. It shows the velocity gradient on both sides
not to be mirror-symmetric (the short white lines mark angles $\pm$30$^\circ$
to the vertical of the diagram). In fact the redshifted component appears to drop
less steeply than its blueshifted counterpart. This would indicate the
{\it rotation velocity} to {\it increase} with distance from the axis (solid
rotation?), which might in principle be used to put lower limits on the magnetic
and frictional forces active in the shell. 

The single-dish observations of CO(2-1), (6-5), and (7-6) complement this
SMA picture. Of course at the velocities of the ambient gas these lines
are optically very thick; it therefore takes the rare isotopomeres to discern
any near-ambient-velocity structure close to the outflow axis. $^{13}$CO(2-1)
shows a clear and simple picture: over a distance of 
at least some 50$''$ along the outflow two ranges of increased
intensity run parallel to the outflow axis, one on either side, of constant
and equal distance ($\le$ 15 $''$) from the axis, and of equal intensity if
observed around v$_{LSR}$ = 7.7 km s$^{-1}$. At lower velocities the ridge to the
North-west begins to dominate, at higher values the south-eastern one. We
thus suspect the  ambient LSR velocity in the
immediate vicinity of the outflow to be around 7.7 km s$^{-1}$, 
with the two ridges marking the edges of a tubular wall that
surrounds the flow axis. The depression between the ridges seems relatively
stronger for the rarer isotopomere C$^{18}$O, maybe indicating preferential destruction
due to less self-shielding against the UV generated in the jet's shock.
A typical $^{13}$CO intensity cross section through the outflow is shown in Fig. 4
for v$_{LSR}$ = 7.7 km s$^{-1}$, taken at downstream distance 45$''$ from the
bending point (marked by the white dot in Fig. 1). This point is from now on to serve as
the zero-point for all single-dish distance determinations along the outflow
because from here the outflow follows a straight line. Its coordinates are
R.A. =  05$^h$35$^m$13.04$^s$, decl.= -05$^\circ$24$'$20.0$''$ (J2000.0). 

\begin{figure}[!ht]
\begin{center}
\includegraphics[scale=1, angle=-90]{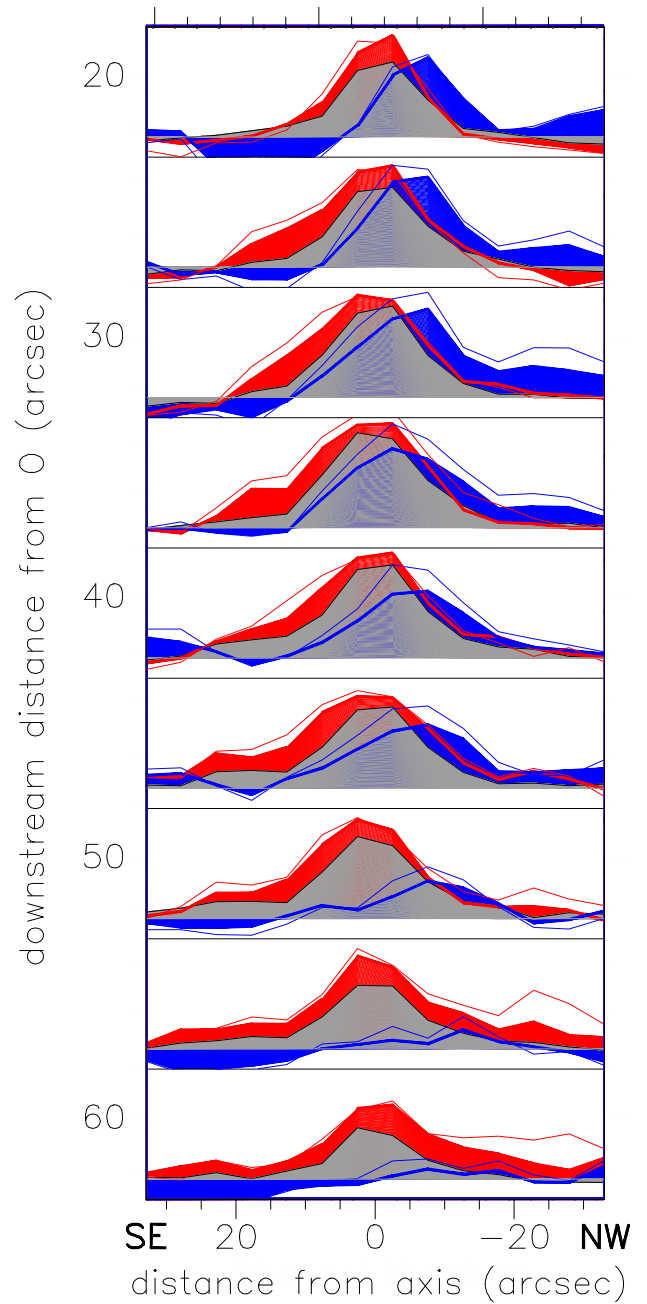}
\caption{\scriptsize  Red-blue asymmetry of CO(6-5) emission across the outflow at increasing
distances from point O. Grey: Cross section at 15 km/s (core of the outflow), red:
at 12.7 km/s, blue: at 4.5 km/s, each integrated over a velocity interval of width 0.5 km s$^{-1}$. 
The T$_{mb}$ scale is -4 to 22 K. The thin colored lines
show the distributions 0.5 km/s closer to the ambient velocity.  }
\label{fig7}
\end{center}
\end{figure}

At velocities a few km s$^{-1}$ away from near-ambient values a structural
asymmetry appears in the emission of both CO(2-1) and $^{13}$CO(2-1), see Fig. 4.
Some 3 to 4 km s$^{-1}$ {\it below} the 7.7 km s$^{-1}$ value the intensity of both species is strongest along
a strip situated between the ambient tubular wall and the outflow axis (as defined by
the high-velocity cross-section profile) on the {\it north-western} side of the axis, at
v$_{LSR}$ some 3 to 4 km s$^{-1}$ {\it above} this value the emission peaks along a corresponding strip on the {\it south-eastern} side.
These two zones are spatially separated by $\sim$ 10$''$ to 15$''$ and by $\sim$ 7 km s$^{-1}$ in radial velocity,
and this asymmetry extends downstream to projected distances of at least 2.5 $\times$ 10$^4$
AU from the source.
The sense of the asymmetry corresponds to our SMA results, but the velocities in question
have a different context: The SMA variations are a superposition of an asymmetry of a few km s$^{-1}$ onto 
the high-velocity components of the outflow, the single-dish data concern motions 
outside the core flow zone.

Typical p-v diagrams taken perpendicularly across the outflow at distance 37$''$
from point O, as obtained with the somewhat higher resolutions of the APEX 
measurements than that of IRAM CO(2-1), are shown in Fig. 5. No important long-range gradient
of the ambient velocities is evident, and in particular there appears no obvious tilt across
the outflow axis (the horizontal white line). However, in both diagrams one notes conspicuous
excursions of the iso-intensity lines, somewhat anti-symmetrically offset from the outflow's
axis, at velocities slightly beyond the optically thick v$_{LSR}$ values, i.e.
below 5 resp. above 11 km s$^{-1}$ (big white circles).
These might seem spurious; but when observing at different downstream
positions one notes their appearance in each p-v crosscut between
20$''$ and 60$''$ from point O (see Fig. 6), contrary to some other local excursions also
visible in Fig. 5 which do not persist across this range.
Fig. 6 does show noticeable differences between the high- and the low-velocity
diagrams.
But complete antisymmetry between the encircled regions of Fig. 5 cannot be
expected because the outflow proper will have different effects on the two
velocity windows, contaminating the redshifted rotation component.
This component seems to extend further from the axis, and it may consist
of two somewhat distinct parts, one at the axial distance of the blueshifted
component (about 5$''$), the other one further away (see panel 37 of Fig. 6).
Higher resolution studies and better knowledge of the central outflow's
contribution are needed to decide this issue.

In the two transition regions between the typical excursion and the ambient
velocities, {\it i.e.} at around 6 resp. 10.5 km s$^{-1}$, there are hints in Fig. 5
(as well as at other downstream distances) of emission {\it deficits} at about
the same offsets from the axis as those where the excursion emission peaks
(Isointensity lines change from convex to concave along the v$_{LSR}$ axis).
This seems to indicate {\it in situ} acceleration in opposite directions on either side
of the outflow, rather than a mere ambient velocity gradient to be at the root of
the observed velocity antisymmetry.

\begin{figure*}[!ht]
\begin{center}
\includegraphics[scale=.9, angle=-90]{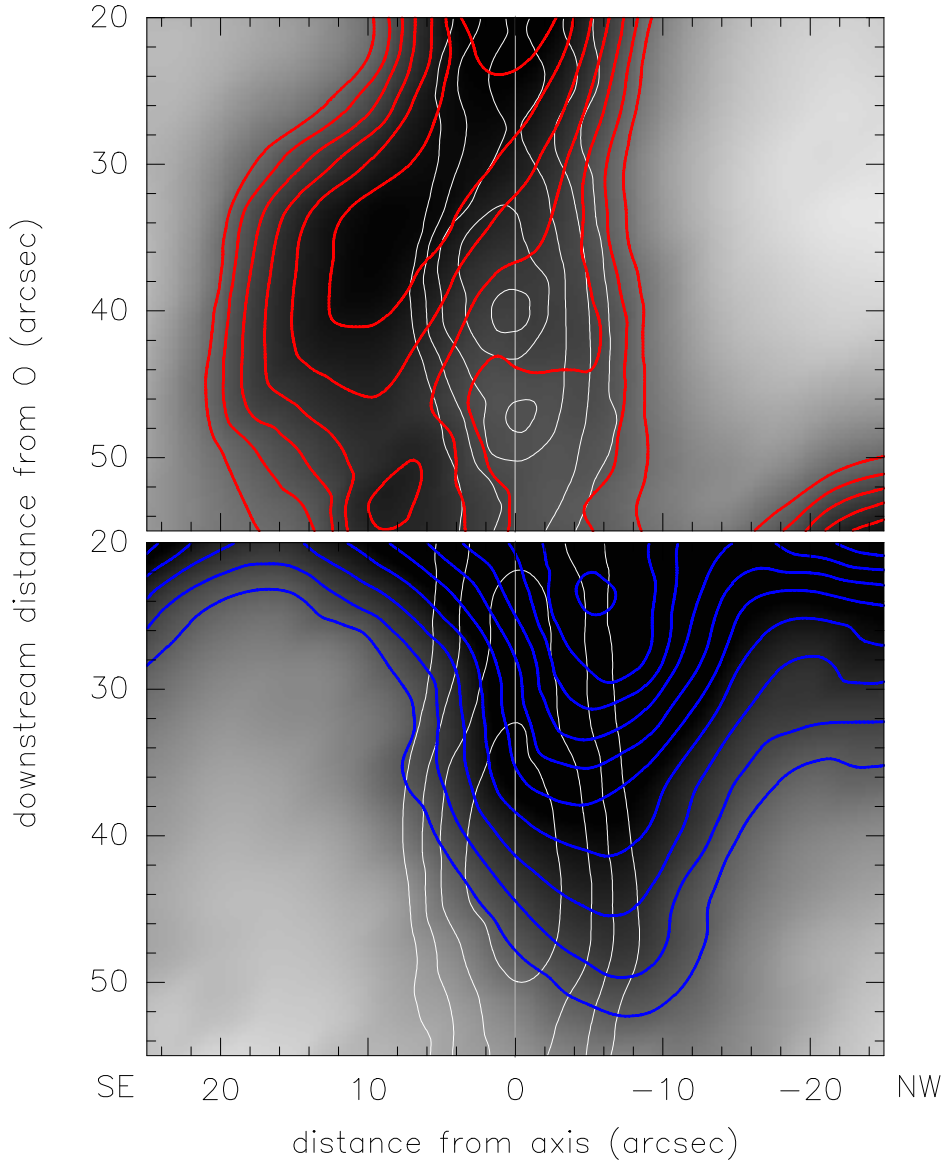}
\caption{\scriptsize CO channel maps (channel width equal to 1.5 km s$^{-1}$) 
centered on velocities 12.0 km s$^{-1}$ (top, CO(7-6)) resp. 4.25 km s$^{-1}$ (bottom, CO(6-5)).
The higher-velocity components of the outflow (here: integrated from 14.5 to 15.1 km s$^{-1}$)
are indicated by the white lines. Note that in the top panel a contribution
from the outflow proper has been cursorily removed (see text).
The coordinate system here is
rotated about point O by -28$^\circ$. }
\label{fig6}
\end{center}
\end{figure*}

Spatial intensity profiles across the flow (Fig. 7), taken for the two velocity excursions in
question at various downstream distances from point O, clearly show a persistent
``red-blue'' asymmetry about the axis that is defined by the high-velocity crosscut
here depicted in grey. Mapping the two ``excursion channels'' of 3.5 to 5.0 km s$^{-1}$
resp. 11.25 to 12.75 km s$^{-1}$ results in the spatial 
distributions of Fig. 8 which represent the considerable velocity difference
of some 6 to 8 km s$^{-1}$ over a distance of around 10-15$''$ or 4-6 $\times$ 10$^3$
AU. Of course this variation does not signal an actual velocity difference of
6 km s$^{-1}$ or so between the two sides, the bulk of the line profiles being hidden
by large optical depth such that only their extreme wings can be seen. The fact that
either side shows predominantly just one of the two wings, the ``red'' or the
``blue'' one, indicates however that the true radial velocity shift across the flow
must be roughly of order at least 0.5 km s$^{-1}$, corresponding to 20 km s$^{-1}$ per pc; 
were it much smaller, then one would measure
about the same wing emission on either side.}
This in turn indicates that the rotation velocity between the inner SMA region
and the annular zone of axial distance R = 10$''$ to 15$''$ does not decrease
much faster than 1/R.

Note that in order to somewhat correct Fig. 8 for the contribution from the outflow proper to the
redshifted intensities we have rather cursorily subtracted part of the
outflow's higher velocities (between 14 and 18 km s$^{-1}$) from the data.

 In Fig. 9 we show the SO(6$_5$-5$_4$) redshifted emission
 of the innermost part of the {\it Ori-S6} outflow as mapped with the
 SMA.  The emission is here integrated over velocities between $+$ 8
 and $+$ 25 km s$^{-1}$.  Our SO observations only detect the
 innermost part of the outflow because of the small primary beam size
 response of the SMA and because the beam was centered to the north of
 the millimeter source 139-409. In Fig. 9, we also mark the primary beam 
 of the SMA.
 The molecular emission is well
 resolved and shows a collimated jet with clumpy morphology that is
 being ejected from that source.  The resolved spatial size of each
 molecular gas bullet or clump is about 1000 to 2000 AU.

\begin{figure*}
\begin{center}
\includegraphics[scale=0.25, angle=0]{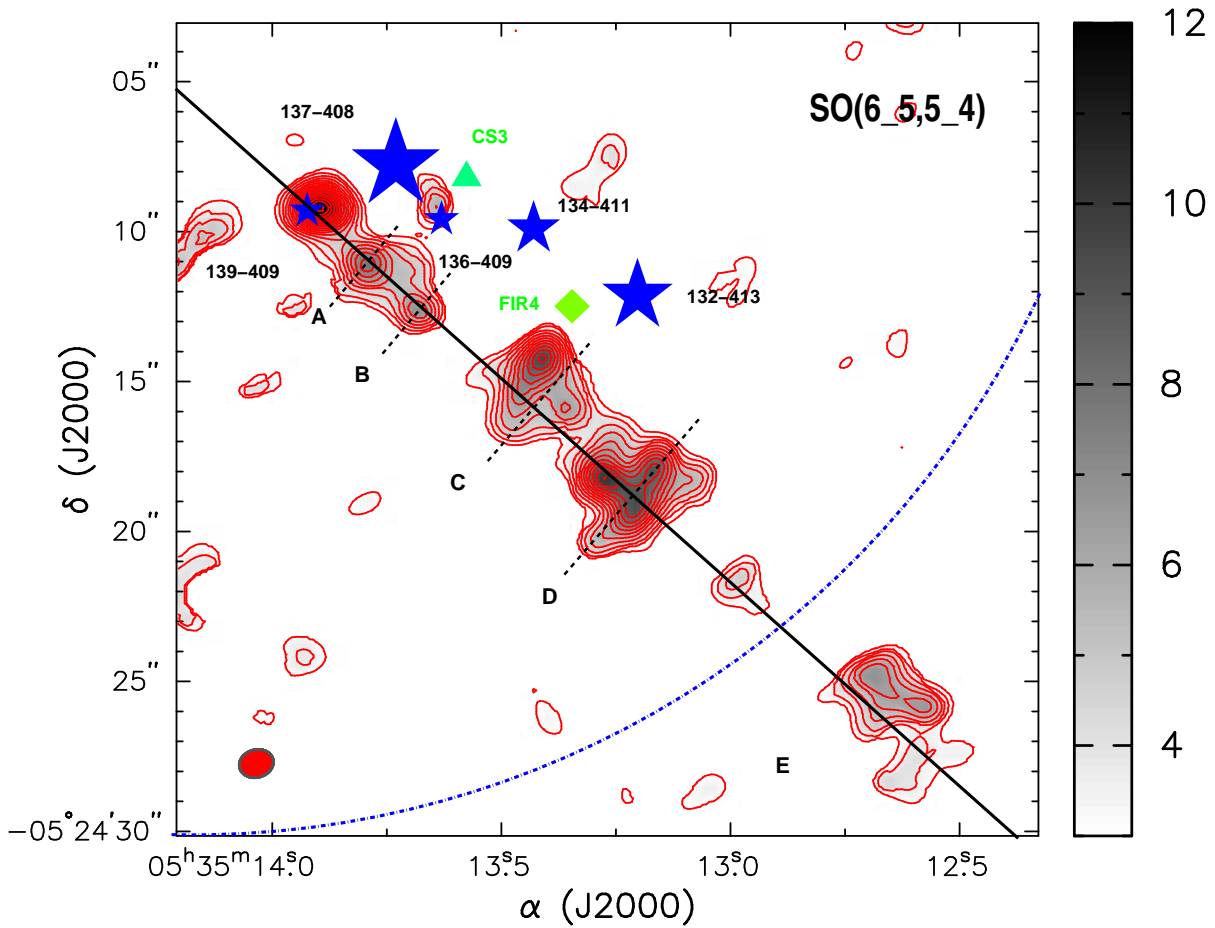}
\caption{\scriptsize SMA SO(6$_5$-5$_4$) moment zero map of the redshifted component 
                     of the {\it Ori-S6} outflow.
                     The contours are -3, 3, 4, 5, 6, 7, 8, 9, 10, 11,
                     12, 13, 14, 15, 16, 17, 18, 19, and 20 times 0.7 
                     Jy beam$^{-1}$ km s$^{-1}$, the rms-noise of the
                     image. The scale bar indicates the integrated
                     molecular emission in units of 1 $\times$ 10$^3$ Jy
                     beam$^{-1}$ km s$^{-1}$. The synthesized beam is
                     shown in the bottom left corner of the image and
                     has a size of 1.13$''$ $\times$ 0.95$''$ with a
                     P.A. of -73$^\circ$. 
                     The molecular emission in the SO map is integrated over 
                     velocities between $+$ 7 and $+$ 25 km s$^{-1}$.
                     The transversal dashed lines
                     across the outflow represent the position of the
                     position-velocity cuts shown in Fig. 10. 
                     The green rhombus and triangle denote
                     the positions of the source FIR 4
                     \citep{Mezgeretal1990} and the millimeter source
                     CS 3 \citep{Mundyetal1986}, respectively. The
                     blue stars mark the position of the millimeter
                     compact sources reported by
                     \citet{Zapataetal2005}. The continuous line
                     traces approximately the symmetry axis of the
                     outflow. The blue arc shows the position and size of the 
                     SMA primary beam. The feature E seems to be
                     no real because it is outside our primary beam.}
\label{fig7}
\end{center}
\end{figure*}

\begin{table}[ht]
\begin{center}
\scriptsize
\caption{Parameters of the SO(6$_5$-5$_4$) Line from the Molecular Gas Bullets}
\begin{tabular}{lccc}
\hline \hline
         & Peak & Half Maximum & LSR Radial \\ & Flux & Velocity Width
         & Velocity \\ Bullet & [Jy Beam$^{-1}$] & [km s$^{-1}$] & [km
         s$^{-1}]$ \\
\hline
\hline
A & 0.18$\pm$0.02 & 11.0$\pm$2.0 & 9.6$\pm$0.7 \\ B & 0.28$\pm$0.02 &
10.0$\pm$1.0 & 12.4$\pm$0.4 \\ C & 0.56$\pm$0.02 & 7.6$\pm$0.4 &
12.1$\pm$0.2 \\ D & 0.34$\pm$0.02 & 10.5$\pm$0.9 & 15.0$\pm$0.4 \\
\hline \hline
\end{tabular}
\end{center}
\end{table}

 Position-velocity diagrams along directions perpendicular to the
 SO jet ({\it i. e.} along P.A. 135$^\circ$) are presented in
 Fig. 10 for four molecular gas bullets (A, B, C and D, see
 Fig. 9).   The bullet named E, seems to not be
 real because it falls far outside the primary beam response of the SMA.
 Three of these bullets (B, C and D) again show
 velocity jumps across the symmetry axis, of the same sense as
 those observed on large scales ( IRAM 30m, APEX, and SMA)  
 but with larger velocity excursions (7 to 11 km s$^{-1}$) over scales 
 of $\sim$ 1000 AU, see
 Table 1 where we give the parameters of the Gaussian least square
 fits to the profiles.  The kinematics of the molecular gas in the
 bullets seems consistent with a rigid body law where the velocity is
 proportional to the distance from the rotation axis.  This feature
 seems least evident in the innermost bullets, perhaps because of our
 poor spectral resolution (4 km s$^{-1}$) as well as their very
 compact size of around one SMA beam ($\sim$ 1$''$).  Note that the molecular bullets
 appear to increase their radial velocities with distance from the
 ejecting object (Table 1).  A similar increase had already, on the
 much larger 30m scale, been discovered out to 70$''$ from the source
 in the CO flow
\citep{Schmid-Burgketal1990}.

 \citet{Zapataetal2007} reported the millimeter source 139-409 to be a
 circumbinary molecular ring, of size a few hundred astronomical units
 ($\rm 294~AU \pm 14~AU \times 207~AU \pm 18$ AU), that is produced by
 two intermediate-mass stars with very compact circumstellar disks of
 sizes and separations less than 50 AU.  The circumbinary disk is seen
 almost edge-on with $\rm PA=87^\circ $.  The redshifted molecular gas
 is located toward the west, the blueshifted one toward the east. Note
 that the sense of rotation of the molecular material in this
 circumbinary disk is opposite to that found in the jet.

\begin{figure*}
\begin{center}
\includegraphics[scale=0.3, angle=-90]{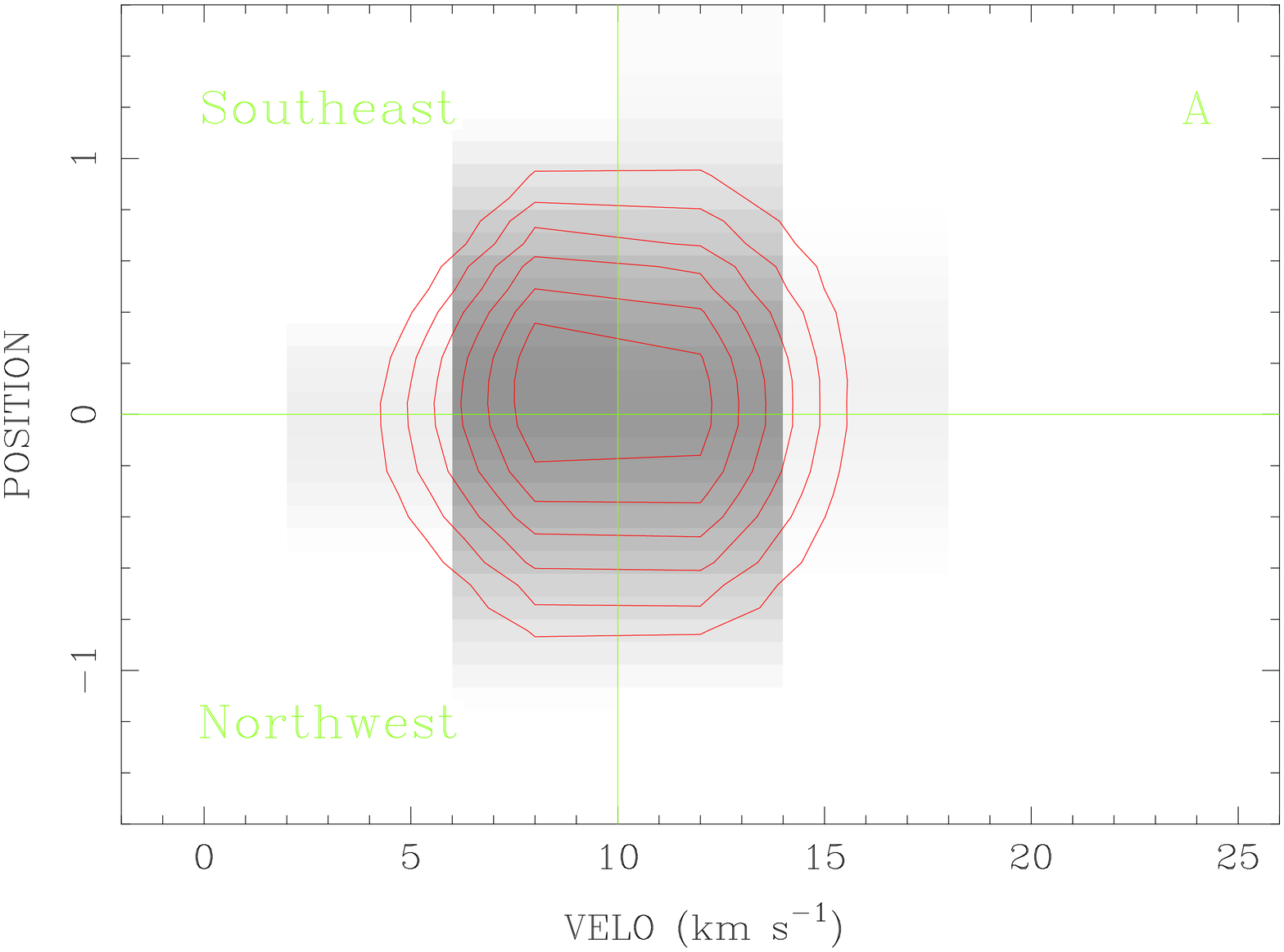}
\includegraphics[scale=0.3, angle=-90]{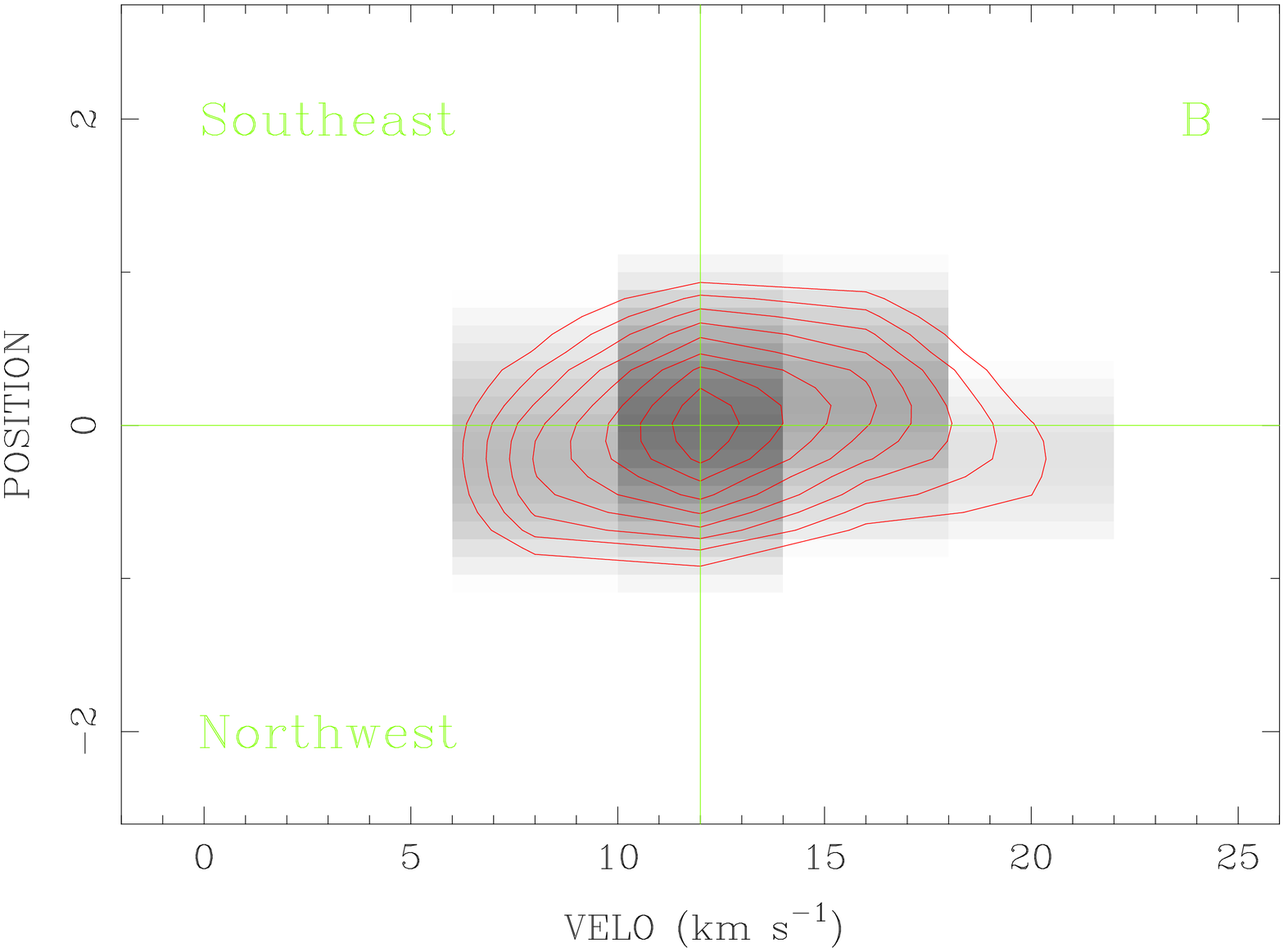}\\
\includegraphics[scale=0.3, angle=-90]{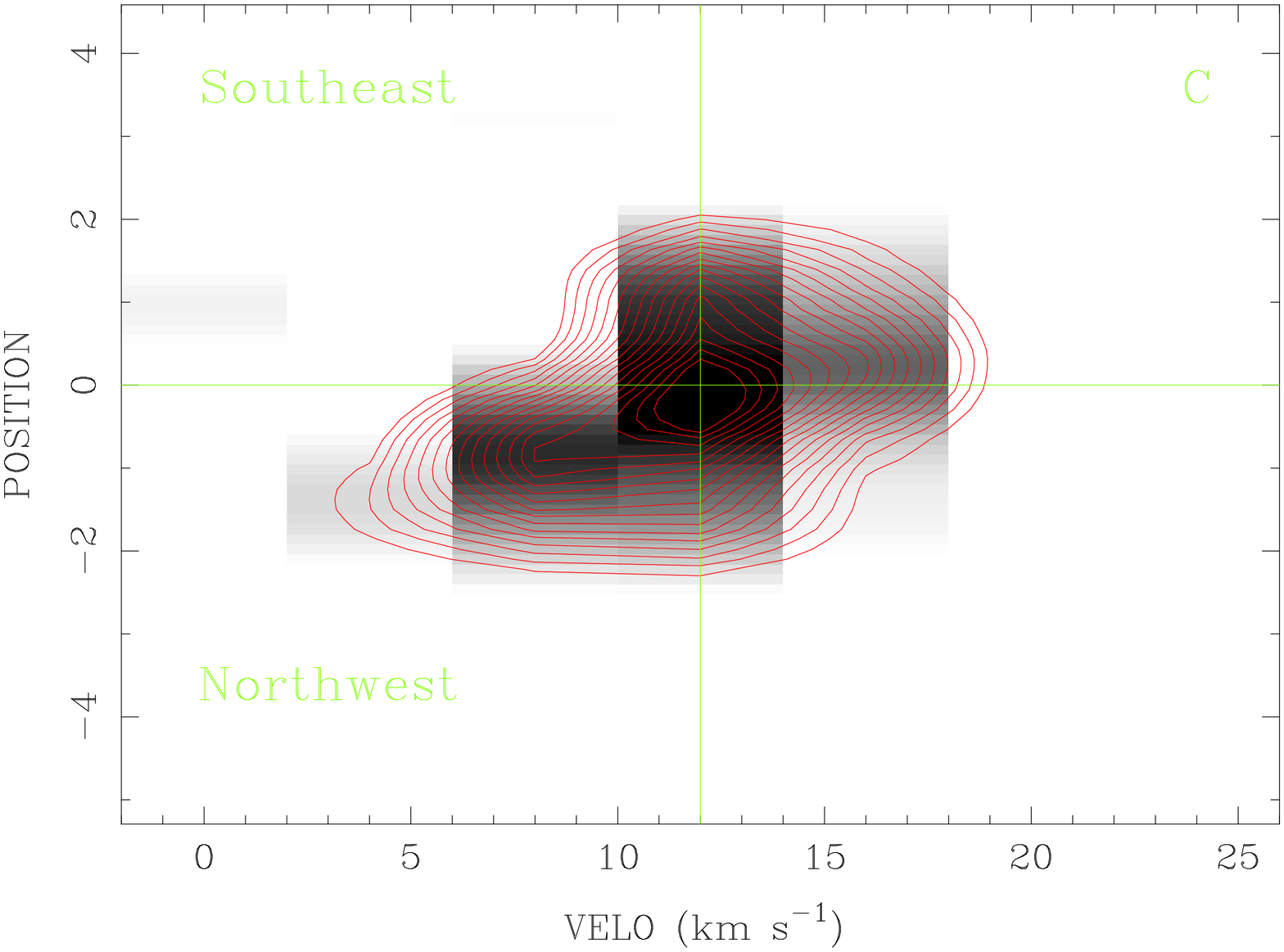}
\includegraphics[scale=0.3, angle=-90]{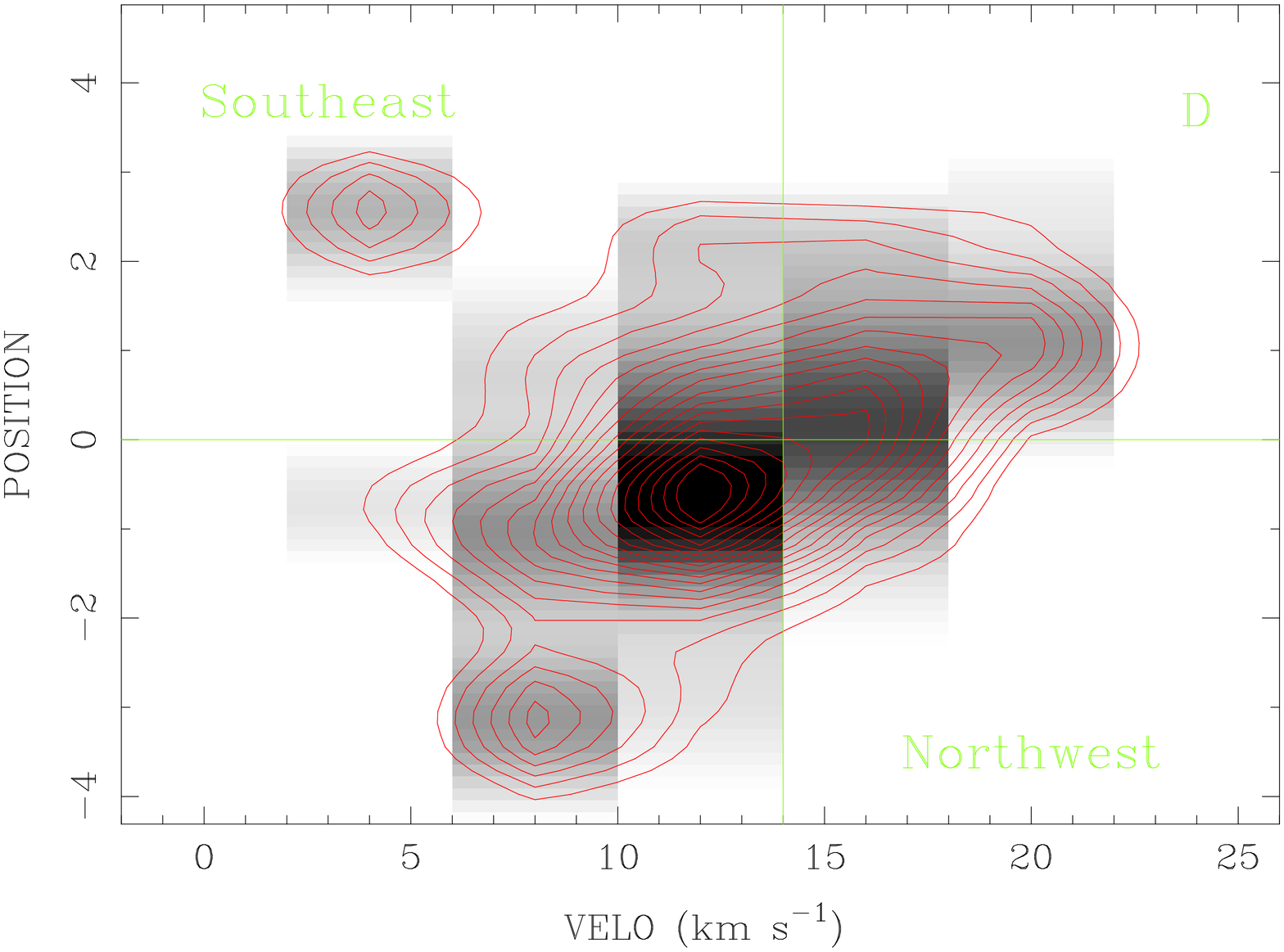}
\caption{\scriptsize Position-velocity diagrams of the four transversal cuts 
 (along a P.A. 135$^\circ$) across the outflow's redshifted component shown in 
Fig. 9. 
 The contours are -3, 3, 4, 5, 6, 7, 8, 9, 10, 11,
 12, 13, 14, 15, 16, 17, 18, 19, 20, 23, 25, 27 and 30 times 20 mJy
 beam$^{-1}$, the rms-noise of the image.  The systemic LSR velocity
 of the ambient molecular cloud here is about 7 to 8 km s$^{-1}$. The green
 lines in each panel mark the position of the symmetry axis of the
 collimated jet and the LSR radial velocities of the molecular bullet
 from Table 1. The velocity resolution is 4 km s$^{-1}$. The
 scale is in arcsec. }
\label{fig8}
\end{center}
\end{figure*} 

\section{Discussion}

Three independent observations all show velocity asymmetries about the
outflow axis that suggest rotation on different length scales. Seen along
the redshifted lobe down from the origin this rotation would be clockwise
for the SO clumps, the CO jet shell, and the ambient envelope alike.

Although this congruence of velocity shifts on three different
length scales lends some credence to a rotation model for Ori-S6 we have to
discuss alternative explanations for the observed red-blue
asymmetries.  The most obvious alternative would
be a general large-scale velocity gradient in the region. Indeed, as the p-v diagram of
Fig. 5 shows, there is some overall velocity
change in the brightest component, from higher to lower velocities along a
direction roughly SE to NW. 
Between the edges of this diagram, {\it i.e.} over a distance of  80$''$, peak v$_{LSR}$ is seen to change by about 1 km s$^{-1}$.
Were this a smooth gradient, it would amount to $\sim$ 5 km s$^{-1}$ per pc, a
value high but not uncommon for molecular cloud clumps. Over the 5$''$
resp. 10$''$ distances between the two sides of the CO jet shell (Fig. 2)
resp. the ``ambient'' CO tube (Fig. 4) this would however amount to velocity
asymmetries below 0.1 km s$^{-1}$, much less than what is observed. 
In fact, over these distances the brightest component does not seem to
vary at all in velocity; only beyond, and then only in the SE part of the diagram
over the very limited extent of some 10$''$, a much larger gradient appears,
of order 60 km s$^{-1}$ per parsec.
This huge gradient connects suspiciously to the velocity dominant still further
away from the flow.
One could postulate such a large, spatially very limited velocity gradient that by chance coincides
with the outflow positions. But that gradient would have to be matched
closely with the flow for these asymmetries to be evident over length scales
of order 60$''$. This seems unlikely.
Also, the convex-concave transitions mentioned above appear to favor {\it in situ}
acceleration over any effects of an ambient velocity gradient.
We thus discard the notion of the
asymmetries being caused by such a gradient in the ambient gas.

Since the outflow originates in a binary system \citep{Zapataetal2007}
a recent calculation
\citep{Murphyetal2008}
of two nearly parallel jets of unequal speed stemming from such a
system may be of relevance for Ori-S6.  In that model the two jets
eventually merge, as witnessed by a persistent kink in the final
structure. Furthermore, due to binarity precession begins to show
after some time in the form of a bending jet trajectory.  Although
neither of these effects can as yet be definitely excluded for our
inner jet, SMA data of the large-distance CO structure clearly speaks
against any sizeable precession or kink. Nor would it seem likely that
the two flow velocities should not over their large common path gradually
adjust to each other and thus wipe out any initial differences in
speed. We therefore look for other explanations for the observed
velocity jumps across the flow axis.

Soker's (2005) alternative proposal, that left-right asymmetries in
jets could result from (non-magnetic) interaction at the base between
jet and a warped disk, would, on the other hand, in our case at the
very least require a warp spatially static over the long times that it
takes to build up the outflow out to 100$''$ from the source: 0.2 pc /
15 km s$^{-1}$ is of order 13,000 years, so even accounting for
possibly reducing projection factors a very large period of standstill
would be demanded of the close binary system at the source of Ori-S6.
Hence we also discard Soker's mechanism and relate the observed jumps
to rotation instead.

 Where would this rotation originate? The CO jet shown in Fig. 2 is
considered ejected from a rotating protostellar disk and then accelerated
and collimated by MHD forces. \citet{Andersonetal2003} provide a formula
that allows to relate jet properties measured at large distances from the disk
to the position (the ``footpoint'') on the disk from where the observed
jet section first emerges:

\begin{equation}
\varpi_{0}\approx0.7\AU\left(\frac{\varpi_{\infty}}
{10\AU}\right)^{2/3}\left(
\frac{v_{\phi,\infty}}{10\kms}\right)^{2/3}
\left(\frac{v_{p,\infty}}{100\kms}\right)^{-4/3}
\left(\frac{\stellarmass}{1\solarmass}\right)^{1/3}
\label{eqn:1}
\end{equation}

Here $\varpi_{\infty}$ is the observed radial distance of the jet shell from the flow
axis, $\varpi_{0}$ the radius on the disk from where that shell's material leaves,
v$_{\phi,\infty}$ and v$_{p,\infty}$  are the toroidal and poloidal velocities observed for the shell
at $\varpi_{\infty}$, and $\stellarmass$ the mass of the (proto)star at the center of the disk. In our case we assume from
observation $\varpi_{\infty}$ $\sim$ 2.5$''$=1000 AU, v$_{\phi,\infty}$ $\sim$ 2 km s$^{-1}$,  v$_{p,\infty}$ 
$\sim$ 10/sin $\alpha$ km s$^{-1}$ and $\stellarmass$ $\sim$ 2 to 5 M$_\odot$, with $\alpha$ the unknown angle 
between flow direction and plane of the sky. This results in a footpoint radius of 140/sin $\alpha$ AU
for the jet component shown in Fig. 2. Since the apparent poloidal flow
velocity of $\sim$ 10 km s$^{-1}$ relative to ambient is relatively low for molecular 
jets it seems likely that $\alpha$ is small. For $\alpha \le$ 30 $^\circ$ the
footpoint radius would drop to below 56 AU, a value that fits the 
above-mentioned disk dimensions of the binary components very well.
Furthermore, similar values for the footpoint radius are found
for the SO(6$_5$-5$_4$) molecular jet.

A similar estimate for the CO envelope of nearly ambient velocity, shown in cross-section in Fig.
5, must await more detailed investigations since its apparent v$_{p,\infty}$ value is
(very) small and not determined; likewise v$_{\phi,\infty}$ can be estimated only very roughly,
as explained above -- the asymmetric wings say little quantitatively about
the bulk toroidal motions of the envelope. A trial with $\varpi_{\infty}$ $\sim$ 5$''$,
v$_{\phi,\infty}$ $\sim$ 0.4 km s$^{-1}$, and v$_{p,\infty}$  $\sim$ 1 km s$^{-1}$ 
would result in footpoint radii about a factor 10 larger than the CO jets shell's,
suggesting an origin quite different from that of the jet's disk.

\citet{Andersonetal2003} also provide an expression for the ratio between toroidal
and poloidal components of the magnetic field strength at the observed jet positions
(see their eq. 2). Using the same parameter values as employed for estimating
the jet shell's footpoint radius and setting $\alpha$ to 30$^\circ$ we arrive at
a field strength ratio (toroidal over poloidal) of around 6. The magnetic
field in the shell thus seems to be tightly wound up, thereby keeping the shell
material well collimated by its hoop stresses.

Where in the system 139-409 the jet actually originates has yet to be
determined.  The sense of rotation of the circumbinary ring is nearly
opposite to that of jet and outflow, and the jet leaves the system
under an angle of 45$^\circ$ with the ring plane. One should therefore
expect the origin of the flow to be at the circumstellar disk of one
of the binary components.  The two disks need not align with the
ring. Further deep observations of highest resolution will be required
to clarify the jet-disk connection.

\section{Summary}

The {\it Ori-S6} outflow presents a promising laboratory for future
studies of magneto-centrifugal models of jet acceleration. We have
observed the compact jet and its larger-scale molecular envelope on three
different spatial scales and found the following:

\begin{itemize}

\item The SO bullets in the jet, its CO shell, 
as well as the more distant CO envelope all show rotation about the outflow axis;
the sense of rotation is the same for all;

\item The inner jet is composed of individual bullets that appear to follow a
solid body rotation law with peak $v_{rot}$ values around 5 km
s$^{-1}$.  The CO shell, at a distance of about $\sim$ 1000 AU from 
the axis, rotates at $v_{rot}$ $\sim$ 2 km s$^{-1}$; 
the CO envelope, at a distance of $\ge$ 2000 AU from the
axis, i.e. several times the bullets' radii, rotates at $v_{rot}$ $\ge$ 0.5
km s$^{-1}$;

\item The rotation is observed out to at least 25,000 AU downstream from the source;

\item  The magnetic field lines embedded in the jet's CO shell can well
thread a protostellar disk of radius $\sim$ 50 AU. For the wider CO envelope
the footpoint in the equatorial plane of the disk must lie considerably further out;

\item The exact identification of the outflow source remains open since at the 
obvious position there is a circumbinary ring containing two
intermediate mass stars. Orientation and sense of rotation of the {\it
ring} do not coincide with those of the outflow.

\end{itemize}

All these topics are amenable to more extended observations that aim
at problems of star formation, such as angular momentum budgets
and high-velocity mass ejection.

\acknowledgements We would like to thank Dr. Bernd Klein
for having taken the recent CO data on the 30m telescope, and
Dr. Rainer Mauersberger for reactivating the IRAM data of 1989.

Facilities: {\it IRAM 30m APEX and SMA}

\bibliographystyle{aa}
\bibliography{biblio}

\end{document}